# A proposal for a two-slit experiment with a mirror

G. Sampath
10 Schalks Crossing Rd #501-233, Plainsboro, NJ 08536

**Summary**.  A modified version of the two-slit experiment is proposed in which the moveable detector/counter used to obtain the fringe distribution by counting single photons at different positions on the screen plane is replaced with a mirror positioned at an angle to the screen plane.  A single photon arriving at the mirror through one of the slits is reflected at one of two angles along one of two divergent paths.  It is then detected by one of two detectors, one on each of the paths, for different positions of the mirror on the screen plane. This modification results in the ability to generate which way information based on the path taken by the reflected photon as well as the distribution of photon counts on the screen plane, one for each slit.  Since the photon does not experience any change other than the reflection at the mirror the question whether fringes are also present because of a priori interference effects when no mirror is present can be answered by examining the two a posteriori distributions. Three possible outcomes are considered and explanations proposed for them. The experiment is shown to be feasible with existing technology and can be performed in most physics laboratories.

## 1  Two-path experiments, which-way information, erasure, and wave-particle duality

The two-slit experiment (and the related one using beam-splitters and mirrors) has long been considered one of the enduring mysteries of physics.  In it photons from a collimated source that are widely separated in time pass through a diaphragm with two closely-spaced narrow slits and strike a photographic screen (or are detected by a moveable photon detector/counter) to form an interference pattern similar to Young's fringes.  Most books on quantum mechanics[1, 2] go into the associated theory, others discuss the many related gedanken experiments (proposed and studied by Einstein, Bohr, Feynman, Wheeler, Wooters and Zurek, and others[3]) but few examine the experimental side in any detail.  A comprehensive summary of the latter aspect can be found in an article on the web[4] that reviews experiments with different kinds of quantum objects (photons, electrons, atoms, molecules) up to the late 1990's.

The occurrence of interference fringes in the two-slit experiment is widely regarded as providing a strong basis for the non-classical concept of wave-particle duality, which in the Copenhagen or complementarity interpretation says that a single fundamental object like an electron or a photon exists as both particle and wave but never both at the same time.  Evidence for this one-or-the-other behavior has been seen in the experimentally observed mutual exclusion between the appearance of fringes and 'which-way' knowledge of the path taken by a single photon (upper/lower slit or transmission/reflection in a beam-splitter).  Further evidence of wave-particle duality is seen in the restoration of the fringe pattern when the 'which-way' information is 'erased'.  Such erasure experiments[5-7] typically involve entangled photon pairs that are generated in atomic cascades or spontaneous parametric down conversion in which one photon in the pair, the signal photon, is sent through the slits (or equivalently the transmitted/reflected path from a beam splitter) while the other, the idler photon, is sent to a time-amplitude converter and detected along with the signal photon using a coincidence circuit.

In recent years, the possibility that the two manifestations are not strictly mutually exclusive has been explored[8, 9].  (Even earlier, a continuous transition from particle to wave was postulated and studied as a function of the distance from the diaphragm to the screen[10].)  Wave and particle properties are usually associated with two experimentally measurable quantities.  The wave property is recognized in the visibility V of the fringes, defined as the ratio $(I_{max} - I_{min})/(I_{max} + I_{min})$, where $I_{max}$ and $I_{min}$ are respectively the maximum and minimum image intensities (or equivalently the photon counts) in the measured fringe pattern.  V is then restricted to the range 0 to 1.  The particle property is recognized in the ability to determine the path taken by the photon, referred to as the distinguishability D, through the use of a which-way marker.  The latter, depending on the photon property used (for example, polarization), may have a continuum of values, usually normalized to the range 0 to 1. Wave and particle behaviors are therefore not manifested by quantum objects like photons as one-or-the-other binaries but as imperfect duals. It has been theorized that this imperfection is constrained by a tradeoff between V and D which is expressed quantitatively in the form of a so-called duality relation $D^2 + V^2 \leq 1$ ($0 \leq V \leq 1$ and $0 \leq D \leq 1$) and appears to be supported by experiment. With other kinds of quantum objects particle fluxes may replace which-way markers, and equivalent measures such as predictability P of the path taken in terms of the measured fluxes have been formulated, leading to similar relationships (such as $P^2 + V^2 \leq 1$).

Here a modified two-slit experiment is proposed in which a moveable mirror controlled by a servo motor can be used along with detectors/counters to record which way information. Only the most basic features of wave-particle duality are considered, other aspects like entangled pairs, non-locality, erasure, and delayed choice[11] are not included. Yet the proposed experiment is significantly different from the conventional two-slit experiment as well as the beam-splitter-based arrangement. Thus the which-way information generated by it consists of two spatial distributions of photon counts, one for each of the two slits. Being information obtained away from the screen plane after reflection, it can be viewed as a posteriori information in which a priori interference patterns that would normally be detected at the screen plane may or may not be embedded. The question whether knowledge of the path (D) is obtained without destroying the interference pattern (V) can then be answered by examining the two distributions. The experiment, which is arguably feasible with available technology, may provide further information on the relationship between D and V and add to the understanding of wave-particle duality.

In the development below, the following conventions are used for symbols: *italics* for apparatus objects, vertical (roman) for scalars, vertical **bold** for vectors, and vertical **bold sequence** for paths (polylines). For example, *S* is a slit and **S** its position, *M* is a mirror, *D* is a detector, s is the distance between two slits *$S_1$* and *$S_2$*, **$S_1$ $M_1$** is the ray from slit 1 to one end of *M*, and **$S_1$ $M_0$ D** is a path from *$S_1$* through the center of *M* to *D*.

## 2  The proposed experimental setup

Figure 1 is a schematic of the experimental arrangement, which requires only minor modifications of a standard two-slit experiment with a photon source (typically a monochromatic laser) that has the capability of delivering single photons widely separated in time. As in the usual setup using photomultipliers or avalanche diodes (rather than photofilm) for photon detection, *P* is the source, *D* the diaphragm with the slits *$S_1$* and *$S_2$*, d the inter-slit separation, s the slit width, and *S* the screen plane (which is parallel to *D*). However, the photon detector/counter that moves on *S* under the control of a stepper motor is replaced with a mirror *M* of width w placed on *S* at an angle and two photon detectors/counters *$D_1$* and *$D_2$* situated at distances $L_1$ and $L_2$ from **$M_0$**, the center of *M*. As seen in the figure the detectors are centered one on each of the two paths taken by a photon coming from one or the other slit and reflected from **$M_0$**. Its detection at *$D_1$* or *$D_2$* effectively identifies the slit through which it would have passed on its way to the screen plane.

## 3 Design and analysis of the proposed experiment

For simplicity it is assumed that the slit width is very small (s << d, L). For simplicity the analysis is done in two dimensions. A single photon traverses one of the following paths: **P$S_1$M$D_1$** and **P$S_2$M$D_2$**, where **M**(x) is a point on the mirror when the latter is centered at (L, x). With a collimated source *P* that is centered on the slits the distances from *P* to *$S_1$* and *$S_2$* are equal. For the configuration shown let $d_1$ and $d_2$ be the distance from *$S_1$* and *$S_2$* to **$M_0$**. Then with $d_1 = \sqrt{((x - d/2)^2 + L^2)}$, $d_2 = \sqrt{((x - d/2)^2 + L^2)}$, L >> $d_1$ and $d_2$, the times of arrival of the photon at the detectors (measured from the two slits) are

$$t_1 = (d_1 + L_1) / c \qquad t_2 = (d_2 + L_2) / c \qquad (1)$$

(c = speed of light = $3 \times 10^{-8}$ m/sec). For sufficiently large $L_1$ and $L_2$ (say 1 or 2 m.) these arrival times are of the order of a few nanoseconds, which is well within the capability of existing signal analyzers.

As discussed below, as long as the width of the mirror and the detector apertures are small the which-way detectors will be able to distinguish successive peaks in the fringe pattern. Each detector can be accurately positioned in the line of the corresponding reflected ray for a given position x of the mirror by closing the other slit and centering and orienting the detector aperture for maximum signal level (measured as the number of clicks in a fixed time window for a given source intensity).

Two factors determine the width of the mirror:

1) The visibility of the fringe pattern, which may go to zero because of low resolution in the detected peak separation. (This assumes that the fringes are there and have not been washed out. That is, fringes are not detected here with the mirror positioned along the screen plane because of low resolution but would be detected in the conventional two-slit experiment without the mirror.)

2) Misdetection: A detector receives a photon from the wrong slit. That is, a photon passing through slit 1 (slit 2) goes to detector 2 (detector 1).

Both difficulties may be avoided with an appropriate choice of parameter values. Owing to the symmetry about x = 0 (the fringe distribution is an even function), the analysis is limited to x ≥ 0.

1) The condition for the visibility to be significantly higher than zero is that the length of the projection of the mirror on the screen plane as determined by the rays $S_1M_1$ and $S_2M_2$ must be less than half the fringe period $F_s$. This follows from the sampling theorem[12], it sets an upper limit on the mirror angle θ and the width of the mirror. The required condition for all x of interest, $0 \leq x \leq x_0$, is met when the projection is considerably less than $F_s/2$ at $x = x_0$ = at least a few (say 2) fringe peaks away from the center, that is, for

$$x_0 > 2 F_s. \qquad (2)$$

Based on this, let the mirror diameter w' be set to a value significantly less than $F_s$, while the mirror angle θ to $S$ is set to give reflections that clear the diaphragm:

$$w' = F_s/7 \qquad \theta = \pi/4 \qquad (3)$$

With L >> d and $L_1 = L_2 = L_0$ the detector separation $|D_1 - D_2|$ is given by

$$L_{12} \approx L_0 (\gamma_1 - \gamma_2) \qquad (4)$$

2) For a given mirror width a photon from $S_1$ ($S_2$) must not have a reflected path that goes through the aperture of $D_2$ ($D_1$). Let $D_{1L}$ and $D_{2R}$ be the left and right end of the aperture of detectors $D_1$ and $D_2$ respectively. Misdetection will not occur over the span of fringes if the reflected paths for the rays $S_1M_1$ and $S_2M_2$ clear $D_{2R}$ and $D_{1L}$ respectively. Thus for any point $p(x)$, $0 \leq x \leq x_0$, on the mirror when the latter is positioned at (L, x), if

$$\delta_1(x) = \text{angle } N_1(x) \, p(x) \, S_1 - \text{angle } N_1(x) \, p(x) \, D_{2R}(x) > 0 \qquad (5a)$$

then a photon through slit $S_1$ will clear detector $D_2$. Similarly if

$$\delta_2(x) = \text{angle } N_2(x) \, p(x) \, S_2 - \text{angle } N_2(x) \, p(x) \, D_{1L}(x) < 0 \qquad (5b)$$

then a photon through slit $S_2$ will clear detector $D_1$.

For detecting the presence of fringes it is sufficient to obtain the distribution over two or three periods, so the minimum separation in Eqn. (5a) is obtained with $x = 3F_s$, $p = M_1$, and mirror half-width $w_1 = |M_1 - M_0|$. Similarly the limiting case for Eqn. (5b) occurs with x = 0, $p = M_2$, and half-width $w_2 = |M_2 - M_0|$. The required mirror half-width for minimum separation is then $w/2 = \min(w'/2, w_1, w_2)$. The separation $L_{12}$ between the detectors can now be obtained for suitable $L_1$ and $L_2$ for each position x of the mirror along S. The x increments must be small enough to satisfy the sampling theorem and large enough to prevent degradation of the visibility V.

A straightforward calculation shows that no misdetection will occur when min (w', $w_1$, $w_2$) = w' for x of interest, $0 \leq x \leq 3F_s$. Thus consider the following typical choice of parameters in a regular 2-slit experiment (all dimensions are metric):

$\lambda = 700 \times 10^{-9}$ meters     d = 100 microns     s = 1 micron     L = 10 cm

For these values the standard quantum mechanical calculation for the wave function probability distribution of the photons along x gives a fringe peak separation

$$F_s \approx \lambda L / d = 0.7 \text{ mm} \qquad (6)$$

Micromirrors have recently been used instead of (slower) switches in optical communication to change the path of an optical signal[13]. They are available in thin film form[14] (metal on silicon, for example) with diameters ranging from tens of microns to tenths of a millimeter (at the latter end they are considered large area micromirrors), thickness of hundreds of Å to a few microns, and roughness of a few Å (r.m.s.) or tens of nm (deviation from mean). The following value is therefore feasible and satisfies Eqn. (3):

$$w' = 0.1 \text{ mm} \qquad (7)$$

The limiting values for $\delta_1$ and $\delta_2$ as given by Equations (5a) and (5b) with x = 0 and x = $3F_s$ = 2.1 mm respectively are

$\delta_2$ (x = 0) → 0 at $w_2 \approx$ 0.268 mm       mirror at (L, 0)       (8a)
$\delta_1$ (x = 2.1 mm) → 0 at $w_1 \approx$ 0.273 mm       mirror at (L, 2.1 mm)       (8b)

The required mirror width to avoid misdetection is then given by

$$w = 2 \min (0.05, 0.268, 0.273) = 0.1 \text{ mm} \qquad (9)$$

which is realizable with a micromirror. The separation between the detectors does not change much with increasing x, so the values for $L_1$ and $L_2$ can be obtained from Equation (4) at x = 0. $D_1$ and $D_2$ need to be at a considerable distance from the mirror to provide reasonable separation between them because of the smallness of $\gamma_1 - \gamma_2$. The aperture is usually set by the width of the collection slit. The following values can be achieved in the laboratory:

$L_1 = L_2 = L_0$ = 5 meters,     a = 1 mm

leading to

$L_{12} \approx$ 5 mm

at the centers of the apertures, which should be enough for detectors with miniature housings or mountings. Incidentally it is only necessary for $L_1$ and $L_2$ to be individually large, they need not be equal. Making them different provides more flexibility in detector separation. Larger, clearer, and spatially less demanding separations between the detectors can be obtained with another level of reflection by interposing mirrors *M'* and *M''* in the paths **S..D₁** and **S..D₂** (inset, Figure 1). The two paths available to a photon are then **PS₁MM'D₁** and **PS₂MM''D₂**.

The detector counts $N_1(x)$ and $N_2(x)$ can now be obtained for a range of x values by moving the mirror on the screen plane, adjusting the positions of *D₁* and *D₂* for each position of the mirror as noted earlier, and doing the count over a fixed time window. (If only fringes are to be detected it is not necessary to use two detectors since the intensity measurements in two-slit experiments are ensemble measurements. Thus it suffices to do the measurement over the desired time window at one position corresponding to $\mathbf{D}_1$ and then repeat the measurement with the detector at $\mathbf{D}_2$ for the same amount of time.) When the experiment is done with the conventional setup, that is, the mirror is replaced with a single detector on the screen plane under the control of a stepper motor, fringes should appear in the distribution of the photon count N(x). With the proposed setup, the number of photons detected is equal to $N_1(x) + N_2(x)$, the sum of the counts measured by $D_1$ and $D_2$. If the fringes are still present then $N_1(x) + N_2(x)$ should be distributed in the same way as N(x).

The usual precautions apply, such as ensuring the absence of opportunistic reflections or scattering through the appropriate use of, for example, light guards and surface treatments.

**4 Three outcomes and possible explanations for them**

The main question of interest is: do the fringes appear simultaneously with which-way information generated in the divergent paths $\mathbf{MD}_1$ and $\mathbf{MD}_2$? A definitive answer can only come from performing the experiment. In anticipation, three possible outcomes from the proposed experiment can be identified:
1) the path traversed by every photon is obtained fully and concurrently with interference fringes;
2) path knowledge and fringe occurrence are strictly mutually exclusive;
3) a combination of 1) and 2) reflecting imperfect duality between wave and particle behavior.

Each of these possibilities is now considered going by the characteristics and results of some which-way experiments done to date[5-7, 9].

*Possibility 1*. Two arguments are commonly advanced to discount such an occurrence.
1) Explanations that view any attempt to find the path taken by a photon as an intrusion on its unconstrained passage before or after the slits[1, 2], which results in the fringes being washed out: these are commonly based on the argument that the resulting disturbance is sufficient to cause the actual fringe resolution obtained to exceed the fringe spacing obtained with unconstrained passage by an amount required by the uncertainty principle[3].
2) Explanations of results from experiments with sources that generate entangled photon pairs with one photon of the pair going towards the slits and the other moving away on a related path[5-7]: these are based on the argument that the disappearance of fringes when which way information is generated is not a result of uncertainty but strictly due to complementarity. (These authors[5-7] go further to assert that complementarity is a more fundamental aspect of nature than uncertainty, although both this claim and the complementarity-based explanation for the disappearance of fringes have been disputed[8, 15].)
In both arguments there is an implicit assumption that the which-way measurement agency (for example, polarizers in the experiment by Walborn et al.[5]) is entangled with the photon, leading to the reduction or 'collapse'[16] of the wave front to an irreversible observation when the photon reaches the detector.

In contrast, in the proposed experiment there is nothing other than the mirror (whose presence merely adds a phase to the probability amplitude) that interacts with the photon on its way to a detector. No attempt is made to measure or change in any way the other properties of the photon, such as its momentum (except for the direction change due to reversal at the mirror) or polarization. In the standard two-slit experiment a photon exhibiting wave-like behavior simultaneously takes the two paths $\mathbf{PS_1M}$ and $\mathbf{PS_2M}$ so that waves with wave functions $E_1(x) = E\, e^{i(kd' - \omega t)}$ and $E_2(x) = E\, e^{i(kd'' - \omega t)}$ (where the wave number $k = 2\pi/\lambda$, $\omega = kv_p$, $v_p$ being the phase velocity, $d' = d_1$, and $d'' = d_2$) interfere at (L, x) with observed intensity given by

$$| E_1(x) + E_2(x)|^2 = 2\,(1 + \cos(k(d_1 - d_2))) \qquad (10)$$

wherein the familiar fringe pattern can be recognized. With the proposed modification, these two waves are reflected at the mirror at (L, x) and lead to, at one of the detectors, say *D1*, a detected intensity given by

$$|E(\mathbf{D}_1)|^2 = 2\,(1 + \cos(k(d_1 - d_2) + 2(\gamma_1 - \gamma_2))) \qquad (11a)$$

which again shows a fringe pattern. (Here it is assumed that $L_2 > L_1$ so that the plane wave from *S2* interferes with that from *S1* before being detected at *D2*. Also the pattern is slightly shifted from the usual one from Equation (10) because of the phase difference $2(\gamma_1 - \gamma_2)$, which, incidentally, is much less than $k(d_2 - d_1)$ for all x except very close to 0.) Similarly

$$|E(\mathbf{D}_2)|^2 = 2(1 + \cos(k(d_2 - d_1) + 2(\gamma_2 - \gamma_1))) \qquad (11b)$$

Whether this is what actually happens remains to be seen. If the fringes do not appear then the cause of the disappearance has to be associated solely with the presence of the mirror M and the resulting reflection. Such a dependence would also strengthen the complementarity-based explanation for the disappearance of fringes[4-7] as uncertainty does not appear to play a role here. (But see Possibility 2 next.)

*Possibility 2*. Any explanation of this occurrence has to be based on the reflections occurring at the mirror. But such an explanation would be in conflict with the well-known occurrence of fringes in beam-splitter experiments in which mirrors are used in both arms of the interferometer. There is also the likelihood that the individual distributions $N_1(x)$ and $N_2(x)$ may show interference effects even though their sum may not.

*Possibility 3*. For this to occur the distinguishability D in the duality relation $D^2 + V^2 \leq 1$ may have to be based on the only new variables introduced, the mirror width w and the mirror angle θ.

**Figures**

Figure 1. Modified two-slit experiment

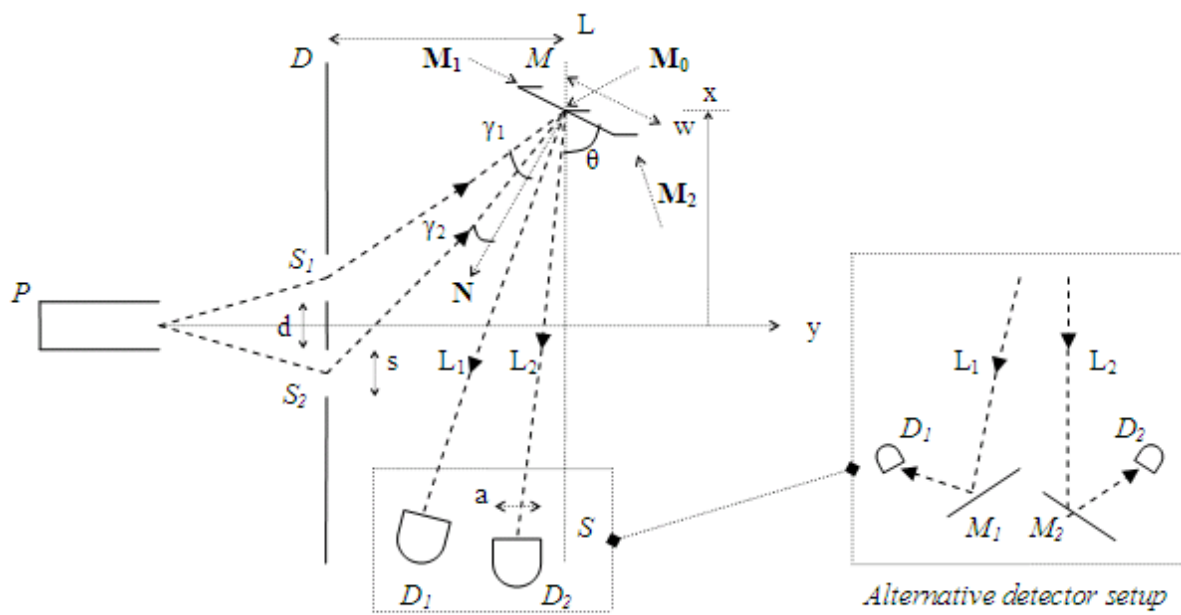

$P$ – Photon source  $D$ – Diaphragm (y=0)  $S$ – Screen plane (y=L)  $S_1$, $S_2$ – Slits
$d$ – Inter-slit distance  $s$ – slit width  $M$ – Mirror  $w$ – mirror width
$N$ – Normal to $M$  $D_1$, $D_2$ – Detectors  $a$ – detector aperture  $x$ – position of $M$ on $S$
$L$ – distance from $D$ to $S$  $L_1$, $L_2$ – Lengths of reflected paths from $M$ to $D_1$, $D_2$
$\gamma_1$ = angle $S_1$-$M_0$-$N$  $\gamma_2$ = angle $S_2$-$M_0$-$N$  $\theta$ = angle made by $M$ with $S$
$M_1$, $M_2$ – micromirrors for wider separation of $D_1$ and $D_2$ (alternative detector setup)

Figure 1. Modified 2-slit experiment